\documentclass[prl,aps,preprintnumbers,superscriptaddress,twocolumn,showpacs]{revtex4}
\usepackage{graphicx,xspace,amsmath,amssymb,amsfonts}
\newcommand{\ket}[1]{\ensuremath{| #1 \rangle}}
\newcommand{\bra}[1]{\ensuremath{\langle #1 |}}
\newcommand{\kb}[2]{\ensuremath{| #1 \rangle \langle #2 |}}
\newcommand{\bk}[2]{\ensuremath{\langle #1 | #2 \rangle}}

\newcommand{\id}{\ensuremath{\mathrm{1}}}

\newcommand{\MI}[1]{\ensuremath{\mathcal{I}(#1)}}
\newcommand{\nn}[1]{\ensuremath{\mathcal{N}_{#1}}}
\newcommand{\Sum}[2]{\ensuremath{\sum_{#1}^{#2}}}

\newcommand{\CK}{Csisz$\acute{\mbox{a}}$r-K$\mathrm{\ddot{o}}$ner (CK)
theorem }

\newcommand{\dt}[1]{\ensuremath{\delta_{#1}}}
\newcommand{\fz}[1]{\ensuremath{f^{z}_{#1}}}
\newcommand{\fy}[1]{\ensuremath{f^{y}_{#1}}}
\newcommand{\fx}[1]{\ensuremath{f^{x}_{#1}}}
\begin{document}

\title{Quantum Tomographic Cryptography with a Semiconductor Single Photon Source}%

\author{Dagomir Kaszlikowski}
\affiliation{Department of Physics, National University of
Singapore, Singapore 117\,542, Singapore}

\author{L. C. Kwek}
\affiliation{National Institute of Education, Nanyang
Technological University, Singapore 639\,798, Singapore}
\affiliation{Department of Physics, National University of
Singapore, Singapore 117\,542, Singapore}

\author{Jenn Yang, Lim}
\affiliation{Department of Physics, National University of
Singapore, Singapore 117\,542, Singapore}

\author{Frederick H.\ Willeboordse}
\affiliation{Department of Physics, National University of
Singapore, Singapore 117\,542, Singapore}

\author{Shiang Yong, Looi}
\affiliation{Department of Physics, National University of
Singapore, Singapore 117\,542, Singapore}

\begin{abstract}
In this paper we analyze the security of the so-called quantum
tomographic cryptography with the source producing entangled
photons via an experimental scheme proposed in Phys. Rev. Lett.\
\textbf{92}, 37903 (2004). We determine the range of the
experimental parameters for which the protocol is secure against
the most general incoherent attacks.
\end{abstract}
 \maketitle
\section{Introduction}

The need to transmit information securely between two parties is
off paramount importance in military and commercial communication.
The discovery of the possibility of secure communication based on
photons has triggered numerous investigations in this field. An
important issue in quantum communication is the ability to
distribute a key securely between distant parties, and a number of
schemes such as the BB84 \cite{bb84} and Ekert91 \cite{e91} have
been proposed for this purpose. In particular, the tomographic
quantum key distribution scheme proposed in Ref.\ \cite{liang} in
which Alice and Bob utilize a tomographically complete set of
observables to distribute the key as well as perform full state
tomography on their key distribution source has been shown to be
powerful as it severely limits Eve's eavesdropping possibilities,
when compared to BB84 or Ekert91. An extension of this scheme to
the more general class of Bell diagonal states has also been
proposed in Ref.\ \cite{bdc}.

On the practical side, quantum key distribution (QKD) is a
sufficiently advanced field so that there is already the
possibility for commercialization of some of the QKD devices.
However, security analysis of generic devices is not always
straightforward.

Recently, using the pulsed optical excitation of a single quantum
dot from a sample of self-assembled InAs quantum dots in a GaAs
matrix, it was shown that it is possible to use linear optics
techniques to induce polarization entanglement between single
photons emitted independently from the source
\cite{yama,yama1,yama2}. Such a technique can be feasibly applied
to produce entangled photons for QKD. In this letter, we exploit
the tomographic QKD scheme \cite{liang} to study the security of
QKD based on such solid state devices.

\section{Tomographic QKD}
In the tomographic QKD scheme \cite{liang}, a central source
distributes entangled qubits to Alice and Bob. Here, we assume
that these qubits arise from polarization-entangled photons
generated using a quantum dot single photon source and the method
described by Fattal et al.\ \cite{yama}. Alice and Bob
independently and randomly choose to measure three tomographically
complete observables $\sigma_x$, $\sigma_y$ and $\sigma_z$ (Pauli
operators) on each qubit. At the end of the transmission, they
publicly announce their choice of observables for each qubit pair.
They then divide their measurement results according to those for
which their measurement bases match, and those for which their
measurement bases do not match. Exchanging a subset of their
measurements allows Alice and Bob to tomographically reconstruct
the density operator of the two-qubit state they share. Those
measurements for which their bases match can be used for key
generation.

According to the discussion presented in Ref. \cite{yama} the
density matrix describing the photon source has the following form
in the (say) $\sigma_z$ basis:
\begin{eqnarray}\label{zstate}
\varrho^{(z)} &=& \frac{1}{2}\left(
\begin{array}{cccc}
2 \alpha &  &  &  \\
 & \beta_1+\beta_2+2\gamma & \beta_1-\beta_2 &  \\
 & \beta_1-\beta_2 & \beta_1+\beta_2-2\gamma &  \\
 & & & 2\alpha
\end{array}\right),
\label{source}
\end{eqnarray}
where
\begin{eqnarray}
\alpha &=& \frac{2g}{\frac{R}{T}+\frac{T}{R}+4g}  \nonumber\\
\beta_1 &=& \frac{\frac{R}{T}+\frac{T}{R}-2V}{2(\frac{R}{T}+\frac{T}{R})+8g}  \nonumber\\
\beta_2 &=& \frac{\frac{R}{T}+\frac{T}{R}+2V}{2(\frac{R}{T}+\frac{T}{R})+8g}  \nonumber\\
\gamma &=&
\frac{\frac{R}{T}-\frac{T}{R}}{2(\frac{R}{T}+\frac{T}{R})+8g}.
\end{eqnarray}
The meaning of the experimentally accessible parameters $R,T,V$
and $g$ is the following: $R (T)$ denotes the reflectivity
(transmittivity) of the beamsplitters in the Mach-Zehnder
interferometer used in the experiment (the ratio $\frac{R}{T}$
reported in Ref. \cite{yama} was $1.1$). The parameter $V$ denotes
the overlap of the wave packets of two consecutive photons
emerging from the quantum dot, and $g$ (denoted in Ref.\cite{yama}
as $g^{(2)}$) is the equal time second-order correlation function.
Furthermore, in order for entanglement to exist in the two photon
state, we require that $V>2g$.

From the point of view of the security analysis, it is more
convenient to express the density matrix (\ref{source}) in the
Bell basis $\{\ket{m_{ab}}\}_{a,b=0,1}$. Here,
$\ket{m_{ab}}=\Sum{k=0}{1}\frac{1}{\sqrt{2}}\omega^{kb}\ket{m_k
m_{k+a}}$ ($\omega = -1$) denotes the Bell state in the $m$th
basis ($m=x,y,z$). We have
\begin{eqnarray}\label{zbell}
\varrho^{(z)}_{\mbox{\tiny Bell}} &=& \left(
\begin{array}{cccc}
 \alpha &  &  &  \\
 & \alpha & &  \\
 & & \beta_1 & \gamma   \\
 & & \gamma & \beta_2
\end{array}\right)
\end{eqnarray}
and in the remaining two other bases
\begin{eqnarray}
\varrho^{(y)}_{\mbox{\tiny Bell}}=\varrho^{(x)}_{\mbox{\tiny Bell}}
&=& \left(
\begin{array}{cccc}
 \alpha &  &  &  \\
 & \beta_1 & & -\gamma \\
 & & \alpha &   \\
 & -\gamma & & \beta_2
\end{array}\right).
\end{eqnarray}

From their state tomography, Alice and Bob can determine how the
parameters $\frac{R}{T}$, $g$ and $V$ affect the security of their
key. From these parameters, they can compute, for each basis, the
maximal strength of correlations between Eve and any one of them.
The \CK \cite{ck} then guarantees that if the correlations between
Alice and Bob are stronger than those between Eve and either of
them, a secure key can be established through one-way error
correcting codes, with the efficiency given by the \emph{CK yield}.
Thus for each basis, there is a CK yield for Alice and Bob's bit
data, and they can find out which basis will give them a positive CK
yield. They will then make use of data only from those bases with a
positive yield to establish their key, rejecting the bits obtained
from the remaining measurements. It is interesting to note that in
the tomographic protocol presented in \cite{liang}, Alice and Bob do
not have to do this as the yield is the same regardless of the
measurement basis.

\section{Eavesdropping}
Suppose we have an eavesdropper Eve in the channel. In order to
make our analysis foolproof, we assume the worst-case scenario in
which she is in full control of the qubit-distributing source, and
that all the factors that contribute to experimental imperfections
(parameters $R,T,g$ and $V$) are due to her eavesdropping
activities.

In order to obtain as much information as possible about the key
generated by Alice and Bob, Eve entangles their qubits with
ancilla states $\ket{e^{(z)}_{ab}}$ in her possession. She prepares the following state:
\begin{eqnarray}\label{zpur}
\ket{\psi_{ABE}} &=&
\sqrt{\alpha}\ket{z_{00}}\ket{e^{(z)}_{00}}+\sqrt{\alpha}\ket{z_{01}}\ket{e^{(z)}_{01}}\nonumber\\
&&{}+\sqrt{\beta_1}\ket{z_{10}}\ket{e^{(z)}_{10}}+\sqrt{\beta_2}\ket{z_{11}}\ket{e^{(z)}_{11}},
\end{eqnarray}
where
\begin{eqnarray}
\bk{e_{a'b'}}{e_{ab}} &=& \left\{%
\begin{array}{ll}
    \dt{aa'}\dt{bb'}, & \hbox{$a=0$,} \\
    \frac{\gamma}{\sqrt{\beta_1\beta_2}}\dt{aa'}(1-\dt{bb'})+\dt{aa'}\dt{bb'}, & \hbox{$a=1$.} \\
\end{array}%
\right.\nonumber\\
\end{eqnarray}
Tracing out Eve gives the mixed state Eq.~(\ref{zbell}) that Alice
and Bob measure and this purification is the most general one as
far as eavesdropping is concerned. Note that because of the
tomography performed by Alice and Bob, Eve cannot prepare a state
that would give her some additional correlations across different
qubits emitted by the source. This considerably reduces the number
of coherent eavesdropping strategies Eve can use. More precisely,
the only possibility of a coherent attack for Eve is to collect
her ancillas and perform some collective measurements on them. In
this paper we do not investigate this scenario and assume that Eve
measures her ancillas one by one.

Eve's purification, when expressed in different bases, reads
\begin{eqnarray}
|\psi_{ABE}\rangle&=&\Sum{k,a=0}{1}\sqrt{\mu_{ak}^{(z)}}\ket{z_k,z_{k+a}}\ket{f_{ak}^{z}}
\nonumber\\
&=&
\Sum{k,a=0}{1}\sqrt{\mu_{ak}^{(y)}}\ket{y_k,y_{k+a}}\ket{f_{ak}^{y}}
\nonumber\\
&=&
\Sum{k,a=0}{1}\sqrt{\mu_{ak}^{(x)}}\ket{x_k,x_{k+a}}\ket{f_{ak}^{x}},
\end{eqnarray}
where
\begin{eqnarray}
\mu_{ak}^{(z)} &=&
\dt{a,0}\alpha+\dt{a,1}\left((\dt{k,0}-\dt{k,1})\gamma+\frac{\beta_1+\beta_2}{2}\right)
\nonumber\\
\mu_{ak}^{(x)}=\mu_{ak}^{(y)} &=&
\dt{a,0}\frac{\alpha+\beta_1}{2}+\dt{a,1}\frac{\alpha+\beta_2}{2}.
\end{eqnarray}
The ancilla kets have the following inner products
\begin{eqnarray}\label{structure}
\bk{\fz{a'k'}}{\fz{ak}}&=& \left\{%
\begin{array}{l}
    \dt{kk'},\hbox{\;\;\;if $a=a'=0$,} \\
    \dt{kk'}+(1-\dt{kk'})\frac{\beta_1-\beta_2}{\sqrt{(\beta_1+\beta_2)^2-4\gamma^2}},\\
    \hbox{\;\;\;\;\;\;\;\;\;\;\;\;\;\;\;\;\;\;\;\;\;\;\;\;\;\;\;\; if $a=a'=1$,} \\
    0,\hbox{\;\;\;if $a\ne a'$.} \\
\end{array}%
\right. \nonumber\\
\bk{\fx{a'k'}}{\fx{ak}}&=&\bk{\fy{a'k'}}{\fy{ak}}\nonumber\\
&=&\left\{%
\begin{array}{l}
    \dt{kk'}+(1-\dt{kk'})\frac{\alpha-\beta_1}{\alpha+\beta_1},\hbox{\;\;\;if $a=a'=0$;} \\
    \dt{kk'}+(1-\dt{kk'})\frac{\alpha-\beta_2}{\alpha+\beta_2},\hbox{\;\;\;if $a=a'=1$;} \\
    -\frac{\gamma}{\sqrt{(\alpha+\beta_1)(\alpha+\beta_2)}}\omega^{k+k'},\hbox{\;\;\;if $a\ne a'$.} \\
\end{array}%
\right.\nonumber\\
\end{eqnarray}

Eve's eavesdropping strategy then proceeds as follows. After Alice
and Bob announce their measurement bases and the basis they intend
to use for key generation, Eve knows which pairs of qubits
contribute to the key and that her ancilla for each of those pairs
is a mixture of four possible states. Formally this can be viewed as
a transmission of information from Alice and Bob to Eve encoded in
the quantum state of Eve's ancilla. To find the optimal
eavesdropping strategy, she has to maximize this information
transfer by a choice of a suitable generalized measurement known as
a Positive Operator Valued Measure (POVM)~\cite{Davies}. For
example, if Alice and Bob chose to obtain their key from the $x$
basis, Eve would obtain the following mixed state of ancillas,
\begin{eqnarray}
\varrho_E^{(x)}&=&\Sum{k,a=0}{1} \mu^{(x)}_{ak}
|f^{x}_{ak}\rangle\langle f^{x}_{ak}|.
\end{eqnarray}
Eve has to find the optimal measurement that will extract from the
transmission as much information as possible, the so-called
\emph{accessible information}.

\section{Incoherent Attack}\label{incoh}
The conditions for which our protocol is secure against Eve's
incoherent eavesdropping attack is given by the CK theorem: a
secure key can be generated from a raw key sequence by means of a
suitably chosen error-correcting code and classical one-way
communication between Alice and Bob if the mutual information
between Alice and Bob ($\MI{A;B}$) exceeds that between Eve and
either one of them (the CK regime). For the protocol considered,
the mutual information between Alice and Eve ($\MI{A;E}$), and Bob
and Eve, are the same so that security is assured as long as
\begin{eqnarray}\label{ckreg}
\MI{A;B} &>& \MI{A;E}.
\end{eqnarray}
Furthermore, the difference in mutual information
$\MI{A;B}-\MI{A;E}$ gives the CK yield for the distilled key. Due
to the asymmetric nature of the state in the $\sigma_z$ and
$\sigma_x$/$\sigma_y$ bases, the yield is different for those
bases. The yield is the same for $\sigma_x$ and $\sigma_y$. As
mentioned earlier, Alice and Bob will choose only those
measurement bases which give them a positive yield and use the
data from those bases for key generation.

We shall now present the POVM that maximizes the information
transmitted from Alice and Bob to Eve for a given basis.

Suppose Eve receives a state in the $\sigma_z$ basis:
\begin{eqnarray}
\varrho_E^{(z)}&=&\Sum{k,a=0}{1} \mu^{(z)}_{ak}
|f^{z}_{ak}\rangle\langle f^{z}_{ak}|,
\end{eqnarray}
where the kets have the structure given by Eq.~(\ref{structure}).
Ancillas from the correlation subspace ($a=0$) are orthogonal to
all other states; those from the anti-correlation subspace ($a=1$)
are in general non-orthogonal among themselves.

In the first step, Eve sorts the mixture of the ancillas into two
sub-ensembles according to the index $a$. This can easily be done
using a projective measurement. After that, depending on the
outcome of the projection ($a=0$ or $a=1$), Eve has a mixture of
two ancilla states each corresponding to Alice and Bob's result.

If she projects into the $a=0$ subspace, Eve will possess a
mixture of equiprobable orthogonal ancilla states
\begin{equation}
\varrho^{(z)}_{E,a=0} =
\frac{1}{2}\kb{\fz{00}}{\fz{00}}+\frac{1}{2}\kb{\fz{01}}{\fz{01}},
\end{equation}
which she can distinguish perfectly.

On the other hand, if she projects into the $a=1$ subspace, she
will obtain a mixture of non-orthogonal ancilla states instead:
\begin{eqnarray}
\varrho^{(z)}_{E,a=1} &=&
\left(\frac{1}{2}+\frac{\gamma}{2\alpha+\beta_1+\beta_2}\right)\kb{\fz{10}}{\fz{10}}\nonumber\\
&&{}+\left(\frac{1}{2}-\frac{\gamma}{2\alpha+\beta_1+\beta_2}\right)\kb{\fz{11}}{\fz{11}}.
\end{eqnarray}
We shall denote the inner product of the two ancilla states by
$\lambda\equiv
\frac{\beta_1-\beta_2}{\sqrt{(\beta_1+\beta_2)^2-4\gamma^2}}$. If
these states are equiprobable, which happens if $\gamma=0$ or $R=T$, the optimal
measurement for Eve would be the so-called \emph{square-root
measurement}~\cite{sqm1, sqm2}. Its POVM is given by
$\{\kb{\omega_0}{\omega_0},\kb{\omega_1}{\omega_1}\}$, where
\begin{eqnarray}
\ket{\omega_{10}} &=&
\frac{1}{1-2\eta}\left(-\sqrt{\eta}\ket{\fz{10}}+\sqrt{1-\eta}\ket{\fz{11}}\right) \nonumber\\
\ket{\omega_{11}} &=&
\frac{1}{1-2\eta}\left(\sqrt{1-\eta}\ket{\fz{10}}-\sqrt{\eta}\ket{\fz{11}}\right), \nonumber\\
\end{eqnarray}
with $\eta=\frac{1}{2}(1+\sqrt{1-\lambda^2})$ being the
probability of determining a given state correctly.

In general, the ancilla states will not occur with the same
probability, and the optimal measurement for Eve will then not be
the square-root measurement. Consider the POVM
$\{\kb{\tilde{\omega}_{10}}{\tilde{\omega}_{10}},\kb{\tilde{\omega}_{11}}{\tilde{\omega}_{11}}\}$,
where
\begin{eqnarray}
\ket{\tilde{\omega}_{10}} &=&
\cos{\theta}\ket{\omega_{10}}-\sin{\theta}\ket{\omega_{11}}\nonumber\\
\ket{\tilde{\omega}_{11}} &=&
\sin{\theta}\ket{\omega_{10}}+\cos{\theta}\ket{\omega_{11}}.
\end{eqnarray}
These states are rotated from the square-root measurement states
by an angle $\theta$. We then have the following conditional
probabilities
\begin{eqnarray}\label{prob}
p\left(\tilde{\omega}_{10}|\fz{10}\right) &=&
\left(\sqrt{\eta}\cos{\theta}-\sqrt{1-\eta}\sin{\theta}\right)^2
\nonumber\\
p\left(\tilde{\omega}_{11}|\fz{10}\right) &=&
\left(\sqrt{\eta}\sin{\theta}+\sqrt{1-\eta}\cos{\theta}\right)^2
\nonumber\\
p\left(\tilde{\omega}_{10}|\fz{11}\right) &=&
\left(\sqrt{1-\eta}\cos{\theta}-\sqrt{\eta}\sin{\theta}\right)^2
\nonumber\\
p\left(\tilde{\omega}_{11}|\fz{11}\right) &=&
\left(\sqrt{1-\eta}\sin{\theta}+\sqrt{\eta}\cos{\theta}\right)^2,
\end{eqnarray}
where, for instance, $p\left(\tilde{\omega}_{11}|\fz{11}\right)$
denotes the probability of getting the result of the measurement
corresponding to the projector
$|\tilde{\omega}_{11}\rangle\langle{\tilde{\omega}_{11}}|$
provided the state $|\fz{11}\rangle$ was sent.

Using the probabilities in Eq.~(\ref{prob}), we can compute the
mutual information between Alice and Eve as a function of
$\theta$. The optimal measurement for Eve is then given by the
$\theta$ that maximizes the mutual information between them.

Suppose now that Eve receives ancillas from the $\sigma_x$ basis. If
Alice measured bit `0' ($k=0$, probability $\frac{1}{2}$), Eve will
obtain the state
\begin{eqnarray}\label{s0}
\varrho_{k=0}^{(x)} &=&
(\alpha+\beta_1)\ket{\fx{00}}\bra{\fx{00}}+(\alpha+\beta_2)\ket{\fx{10}}\bra{\fx{10}},
\end{eqnarray}
and if Alice measured `1' ($k=1$, probability $\frac{1}{2}$), Eve
will obtain the state
\begin{eqnarray}\label{s1}
\varrho_{k=1}^{(x)} &=&
(\alpha+\beta_1)\ket{\fx{01}}\bra{\fx{01}}+(\alpha+\beta_2)\ket{\fx{11}}\bra{\fx{11}}.
\end{eqnarray}
The structure of the (normalized) ancillas is given by
Eq.~(\ref{structure}):
\begin{eqnarray}
\bk{\fx{00}}{\fx{01}}&=&\frac{\alpha-\beta_1}{\alpha+\beta_1}\equiv \lambda_0\nonumber\\
\bk{\fx{10}}{\fx{11}}&=&\frac{\alpha-\beta_2}{\alpha+\beta_2}\equiv \lambda_1\nonumber\\
\bk{\fx{0k'}}{\fx{1k}}&=&-\frac{\gamma}{\sqrt{(\alpha+\beta_1)(\alpha+\beta_2)}}\omega^{k+k'}\equiv\mu\omega^{k+k'}.\nonumber\\
\end{eqnarray}

Now, the \emph{total} state describing Eve's ancillas is given by
\begin{eqnarray}
\varrho^{(x)} &=&
\frac{\alpha+\beta_1}{2}\ket{\fx{00}}\bra{\fx{00}}+\frac{\alpha+\beta_1}{2}\ket{\fx{01}}\bra{\fx{01}}\nonumber\\
&&{}+\frac{\alpha+\beta_2}{2}\ket{\fx{10}}\bra{\fx{10}}+\frac{\alpha+\beta_2}{2}\ket{\fx{11}}\bra{\fx{11}},\nonumber\\
\end{eqnarray}
which has the following eigenkets
\begin{eqnarray}
\ket{g_0} &=&
\frac{1}{\nn{0}}\left(\ket{\fx{00}}+\ket{\fx{01}}\right)
\nonumber\\
\ket{g_1} &=&
\frac{1}{\nn{1}}\left(\ket{\fx{10}}+\ket{\fx{11}}\right)
\nonumber\\
\ket{g_2} &=&
\frac{1}{\nn{2}}\left(\kappa_+\left(\ket{\fx{00}}-\ket{\fx{01}}\right)+
\eta\left(\ket{\fx{10}}-\ket{\fx{11}}\right)\right)
\nonumber\\
\ket{g_3} &=&
\frac{1}{\nn{3}}\left(\kappa_-\left(\ket{\fx{00}}-\ket{\fx{01}}\right) +
\eta\left(\ket{\fx{10}}-\ket{\fx{11}}\right)\right),
\nonumber\\
\end{eqnarray}
where
\begin{eqnarray}
\kappa_{\pm} &=&
\beta_2-\beta_1\pm\sqrt{(\beta_2-\beta_1)^2+4\gamma^2}
\nonumber\\
\eta &=& 2\gamma\sqrt{\frac{\alpha+\beta_2}{\alpha+\beta_1}}.
\end{eqnarray}
The normalization constants $\nn{k}$ ($k=0,1,2,3$) read:
\begin{eqnarray}
\nn{0} &=& \sqrt{2(1+\lambda_0)}  \nonumber\\
\nn{1} &=& \sqrt{2(1+\lambda_1)}  \nonumber\\
\nn{2} &=& \sqrt{\frac{4}{\alpha+\beta_1}\left(\beta_1\kappa_+^2 + 4\gamma^2\left(\beta_1-\sqrt{(\beta_2-\beta_1)^2+4\gamma^2}\right)\right)}  \nonumber\\
\nn{3} &=& \sqrt{\frac{4}{\alpha+\beta_1}\left(\beta_1\kappa_-^2 +
4\gamma^2\left(\beta_1+\sqrt{(\beta_2-\beta_1)^2+4\gamma^2}\right)\right)}.\nonumber\\
\end{eqnarray}

If we adopt $\{\ket{g_0},\ket{g_1},\ket{g_2},\ket{g_3}\}$ as an
orthonormal basis, the optimal measurement for Eve can then be
expressed as
$\{\kb{\omega_0}{\omega_0},\kb{\omega_1}{\omega_1},\kb{\omega_2}{\omega_2},\kb{\omega_3}{\omega_3}\}$,
where
\begin{eqnarray}
&&\left(\ket{\omega_0},\ket{\omega_1},\ket{\omega_2},\ket{\omega_3}\right)
\nonumber\\
&&=\left(\ket{g_0},\ket{g_1},\ket{g_2},\ket{g_3}\right)\left(
      \begin{array}{cccc}
        -a & a & b & b \\
        b & -b & a & a \\
        c & c & -d & d \\
        d & d & c & -c \\
      \end{array}
    \right),
\end{eqnarray}
and in which $a,b,c,d$ are real numbers. These parameters are also
related by
\begin{eqnarray}
a^2+b^2 &=& \frac{1}{2}  \nonumber\\
c^2+d^2 &=& \frac{1}{2}
\end{eqnarray}
so that the operators decompose the identity.

As before, we can compute the mutual information between Alice and
Eve $\MI{A;E}$ for this basis and maximize it over the two
independent variables $a$ and $c$ to obtain the maximum information
that Eve can obtain about Alice's measurements.

It should be mentioned here that the optimality of the above POVMs
for the three bases was deduced and confirmed numerically using the
algorithms presented in Refs. \cite{Fri, Rehacek}.

\begin{table*}[!t]
\begin{tabular}{|c|c|c||c|c|c||c|c|c||c|}
  \hline
{} & {} & {} & \multicolumn{3}{c||}{$\sigma_z$} &
\multicolumn{3}{c||}{$\sigma_{x,y}$} & {} \\
\cline{4-9}
 $\frac{R}{T}$ & $g$ & $V$ & $\MI{A;B}$ & max $\MI{A;E}$ &
yield, $\Delta_{z}$
& $\MI{A;B}$ & max $\MI{A;E}$ & yield, $\Delta_{x,y}$ & overall yield \\
\hline\hline
   1.1 & 0.1 & 0.6 & 0.3478 & 0.6070 & -0.2592 ($\times$) & 0.1872 & 0.4320 & -0.2448 ($\times$) & 0 \\
  {} & 0.02 & 0.4 & 0.7598 & 0.7550 & 0.0048 & 0.1085 & 0.1088 & -0.0003 ($\times$)& $\frac{1}{3} \Delta_{z} = 0.0016 $ \\
  {} & 0.1 & 0.84 & 0.3478 & 0.3528 & -0.005 ($\times$) & 0.3869 & 0.3755 & 0.0114 & $\frac{1}{3} \Delta_{y}+\frac{1}{3} \Delta_{x} = 0.0076$ \\
  {} & 0.1 & 0.9 & 0.3478 & 0.2845 & 0.0633 & 0.4525 & 0.3321 & 0.1204 & $\frac{1}{3} \Delta_{z} + \frac{1}{3} \Delta_{y}+\frac{1}{3} \Delta_{x} = 0.1014 $ \\
  \hline
\end{tabular}
\caption{Table of yields in the three bases for $\frac{R}{T}=1.1$,
and different values of $g$ and $V$. Due to the asymmetric nature
of the state in the $\sigma_z$ and $\sigma_x$/$\sigma_y$ bases,
the yield is different for those bases. The yield is the same in
$\sigma_x$ and $\sigma_y$. Crosses mean that the data from this
basis is not used for the key generation.} \label{tab1}
\end{table*}

Table~\ref{tab1} summarizes the results of the computations for
various values of $g$ and $V$. We fixed the ratio
$\frac{R}{T}=1.1$ (the value reported in Ref.\ \cite{yama}). We
see that for certain values of $g$ and $V$ (the first row of the
table), for which the state is still entangled, the CK yield
(denoted as $\Delta$) is negative in all the measurement bases.
For such states, according to the CK theorem, one cannot extract
secure bits by means of one-way communication (because the CK
yield is zero). More interesting are cases where the CK yield is
negative in one measurement basis and positive in another. In such
cases, Alice and Bob reject the data obtained by measurements in
the basis with negative yield and process only the data from the
basis for which the CK yield is positive. The total CK yield is
then the equally-weighted average of only the positive yields from
the three measurement bases. In the case where all the CK yields
are positive, Alice and Bob use the data from all the bases.

\begin{widetext}
\begin{figure*}[t]
\centering\rule{0pt}{4pt}\par
\includegraphics[width=20cm]{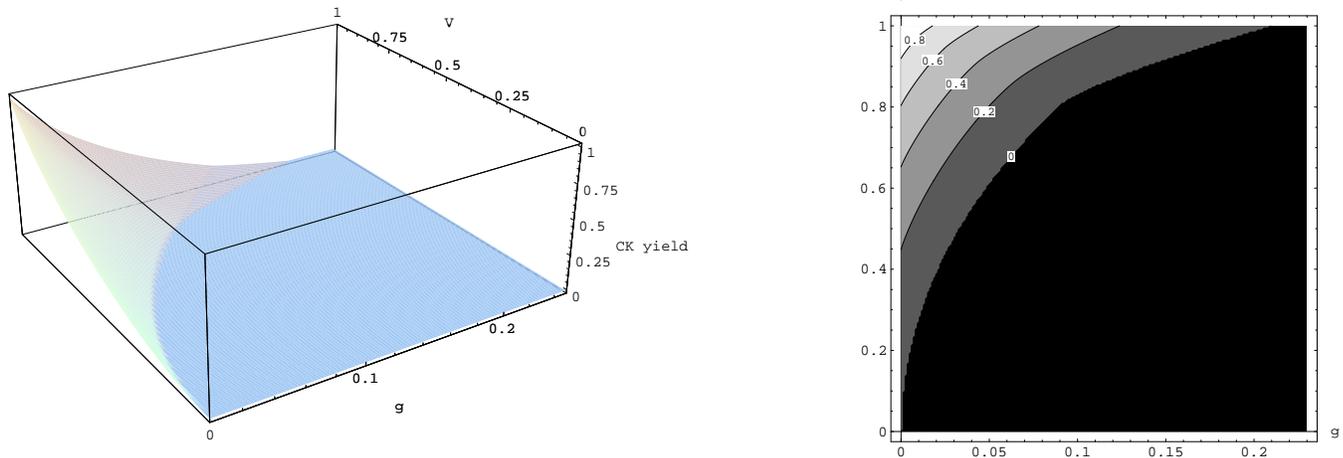}
\caption{Three-dimensional plot of the CK yield for perfect
beamsplitters $R=T$ (\emph{left}), and the corresponding contour
plot (\emph{right}). The threshold for security is given by the
contour for which the CK yield is zero. For $g=0$, security is
guaranteed as long as $V$ is greater than zero, although fewer
secure bits can be distilled for smaller $V$.} \label{fig1}
\end{figure*}
\end{widetext}

In Fig.~\ref{fig1}, we show the overall CK yield plotted against
$g$ and $V$ in the case of perfect beamsplitters ($R=T$). In this
case the state produced by the source is a mixture of Bell states.
The security thresholds for such states have been analyzed in
Ref.\ \cite{bdc}. We observe that the protocol is always secure
against incoherent attacks as long as $g=0$ and $V>0$, although
fewer secure bits can be distilled for smaller $V$. If $V$ is zero
the state becomes separable and, of course, one cannot extract any
secure bits. More detailed analysis reveals that these states (for
which $R=T$, $g=0$ and $V>0$)  have the interesting property that
the mutual information between Alice and Eve is always zero when
Alice performs measurements in $\sigma_x$ or $\sigma_y$ basis.
This is due to the fact that Eve's ancillas corresponding to
different outcomes of Alice's measurements in the $\sigma_x$ and
$\sigma_y$ bases are the same, which means that they do not carry
any information whatsoever about Alice's and Bob's correlations.
Therefore, if Alice and Bob agree on using only the data from the
$\sigma_x$ and $\sigma_y$ measurements (this reduces the
efficiency), the protocol becomes secure against all possible
attacks by Eve (unconditional security). In realistic situations
however, the value of $g$ can be small but not exactly zero (for
example, the value of $g$ reported in \cite{yama1} was 0.02). In
this case, the protocol is secure over a smaller range of $V$.
Even then, we conjecture that the information that Eve can extract
from her ancillas in the $\sigma_x$ or $\sigma_y$ basis is
negligible, and the protocol remains pretty robust against all
possible attacks by her in those bases.

\section{Noisy Channel}
So far, we have excluded the effects of noise in the channel so
that Alice and Bob expect to receive the state `as-is' from the
source. In reality however, this is not the case: Alice and Bob
can expect their quantum channel to be affected by interaction
with the environment. We next consider what happens when there is
symmetric white noise present in the channel, i.e.\ the state that
Alice and Bob expect to receive is of the form:
\begin{eqnarray}
\varrho^{(z)} &=& \frac{1-F}{2}\left(
\begin{array}{cccc}
2 \alpha &  &  &  \\
 & \beta_1+\beta_2+2\gamma & \beta_1-\beta_2 &  \\
 & \beta_1-\beta_2 & \beta_1+\beta_2-2\gamma &  \\
 & & & 2\alpha
\end{array}\right)\nonumber\\
&&{}+\frac{F}{4}\id\otimes\id,
\end{eqnarray}
where we have a proportion $F$ ($0\le F\le 1$) of unbiased noise
admixed to the original state from the source. Analysis shows that
the optimal POVM for Eve is of the same form as that presented
earlier.

\begin{widetext}
\begin{figure*}[!t]
\centering\rule{0pt}{4pt}\par
\includegraphics[width=10cm]{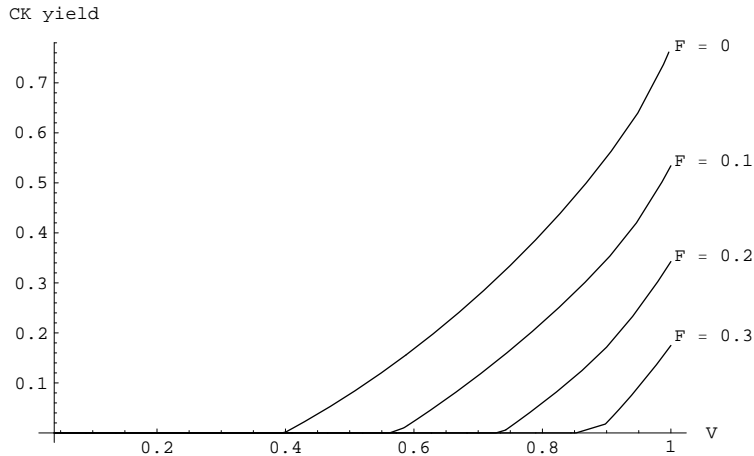}
\caption{Overall CK yield for $\frac{R}{T}=1.1$ and $g=0.02$ and
different amounts of noise in the channel $F$. For a noiseless
channel ($F=0$), when $V\lesssim 0.39$, one can no longer extract
secure bits by means of one-way communication because the CK yield
is zero. As the amount of noise increases, the CK yield drops
until for $F\gtrsim 0.42$, where we will not be able to distill
any secure bits at all (because the CK yield is 0 for all $V$).}
\label{fig2}
\end{figure*}
\end{widetext}

As before, we can obtain the condition for security and the CK yield
for various proportions of noise and this is shown in
Fig.~\ref{fig2}, for fixed values of $\frac{R}{T}=1.1$ and $g=0.02$
(values reported in Ref.\ \cite{yama,yama1}). We can distill less
secure bits as the amount of noise increases.

\section{Conclusion}

We have analyzed the security of the tomographic QKD protocol
using a source of entangled photons produced in the experimental
scheme proposed by Fattal {\it et. al} \cite{yama} against the
most general incoherent attacks.

Using the analysis presented in this paper we can give the number of
secure bits that can be distilled by means of one-way communication
between Alice and Bob as a function of the experimentally accessible
parameters $R,T,g$ and $V$, and for different degrees of unbiased
noise in the channel $F$.

\section{Acknowledgements}
DK and LCK wish to acknowledge support from A*STAR Grant
R-144-000-071-305. DK wishes to acknowledge NUS Grant
R-144-000-089-112. DK and JYL also wish to thank Marek Zukowski
for valuable discussions.


\end{document}